\documentclass{desyproc}

\pdfoutput=1

\begin{document}
%------------------------------------

\title{Top Charge Asymmetry -- Theory Status Fall 2013}

\author{{\slshape Susanne Westhoff}\\[1ex]
PITTsburgh Particle physics, Astrophysics \& Cosmology Center (PITT PACC)\footnote{preprint PITT-PACC-1310}\\
Department of Physics and Astronomy, University of Pittsburgh, Pittsburgh, PA 15260, USA}

\maketitle

\begin{abstract}
I review the current status of the charge asymmetry in hadronic top-quark pair production from a theory perspective. The standard-model predictions for the observables at the Tevatron and LHC are being discussed, as well as possible explanations of the Tevatron excess in terms of new physics. I give an outlook for future investigations of the top-quark charge asymmetry, focussing on novel observables at the LHC.
\end{abstract}

\section{Introduction}
The charge asymmetry in top-antitop production provides us with a test of Quantum Chromodynamics (QCD) beyond leading-order (LO) interactions. It manifests itself in differing angular distributions of top and antitop quarks,
\begin{eqnarray}
A_C = \frac{\sigma_A}{\sigma_S}, \qquad \sigma_{S,A} = \int_0^1\text{d}\cos\theta\left(\frac{\text{d}\sigma_{t\bar t}}{\text{d}\cos\theta}\pm \frac{\text{d}\sigma_{\bar t t}}{\text{d}\cos\theta}\right),
\end{eqnarray}
where $\theta$ is the scattering angle of the top quark ($\sigma_{t\bar t}$) or antitop quark ($\sigma_{\bar t t}$) off of the incident quark in the parton center-of-mass frame. Experimentally, the charge asymmetry is measured in terms of top-antitop rapidity differences,
\begin{eqnarray}
A_C^{\text{exp}} = \frac{\sigma(\Delta y > 0) - \sigma(\Delta y < 0)}{\sigma(\Delta y > 0) + \sigma(\Delta y < 0)}\,.
\end{eqnarray}
In proton-antiproton collisions, the total charge asymmetry is closely related to a top-quark forward-backward asymmetry in the laboratory system, which is measured through the rapidity difference $\Delta y = y_t - y_{\bar t}$ (yielding $A_C^{\text{exp}} = A_C^y = A_C$). In proton-proton collisions, the charge asymmetry induces a forward-central asymmetry, which is measured through the difference of absolute rapidities $\Delta y = |y_t| - |y_{\bar t}|$ (yielding $A_C^{\text{exp}} = A_C^{|y|} \ll A_C$). The sensitivity of $A_C^{|y|}$ to the partonic charge asymmetry is reduced due to $|y_t|-|y_{\bar t}|$ not being invariant under boosts along the beam axis. At the LHC, $A_C^{|y|}$ is further suppressed by a large background from symmetric gluon-gluon initial states.

\begin{figure}[tb]
\centerline{
\includegraphics[width=0.51\textwidth]{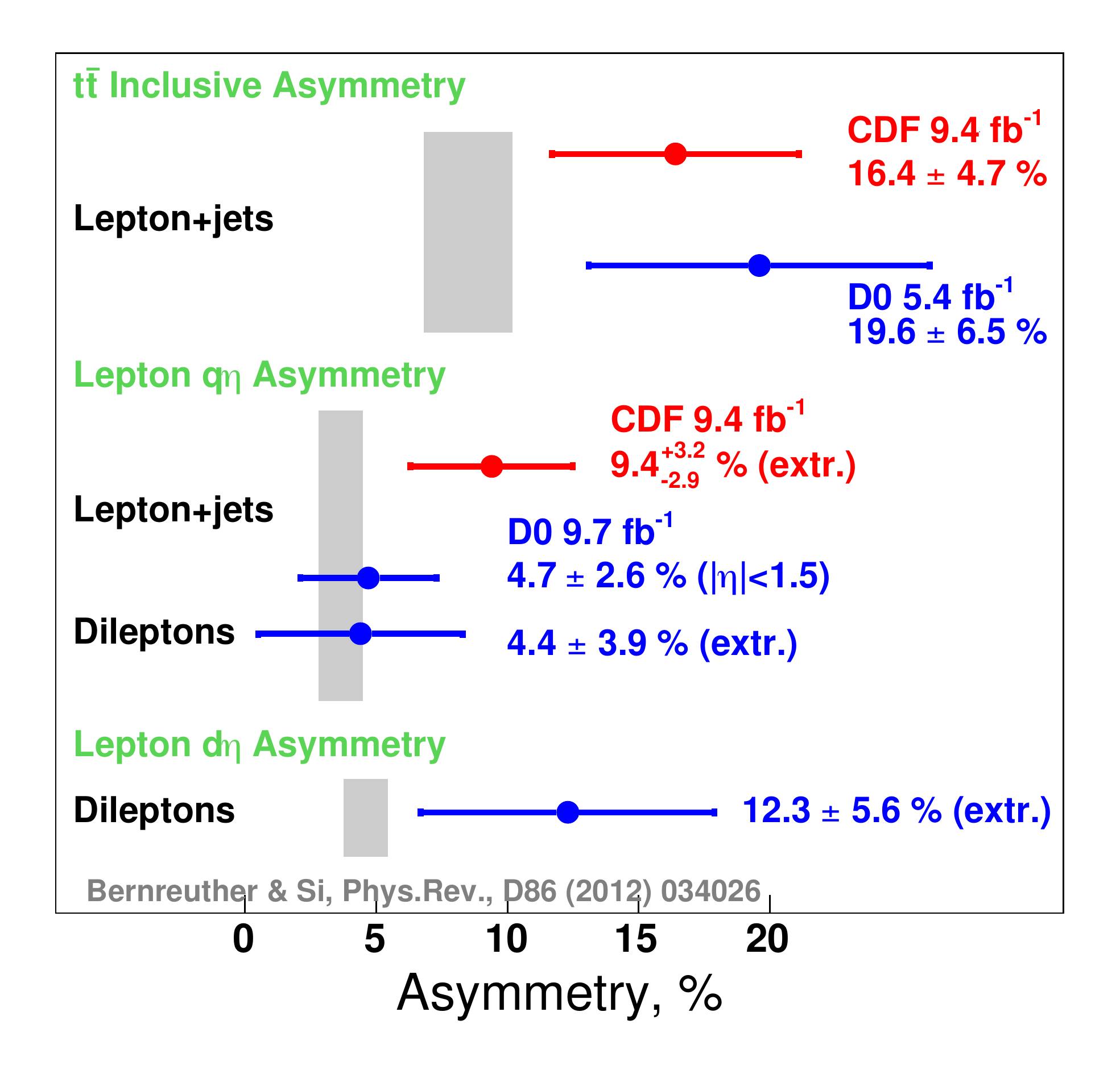}
\includegraphics[width=0.51\textwidth]{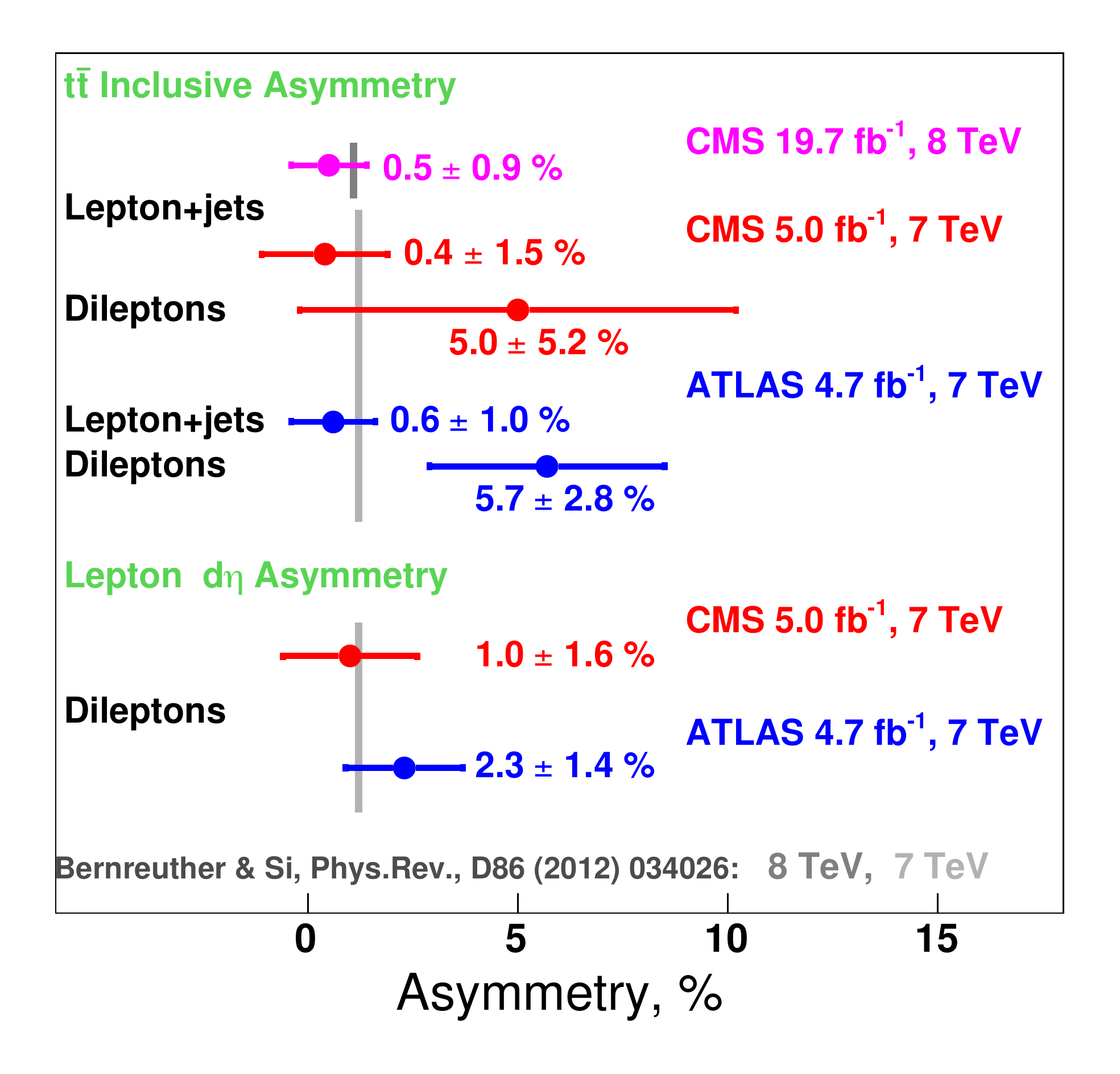}
}
\caption{Charge asymmetry measurements at the Tevatron (left panel) and the LHC (right panel). Shown are the inclusive $t\bar t$ asymmetries $A_C^y$ and $A_C^{|y|}$ in lepton+jets and dilepton final states, as well as the lepton asymmetries $A_C^{\ell}$ ($q\eta$, Tevatron) and $A_C^{\ell\ell}$ ($\text{d}\eta$), defined in Eqs.~\ref{eq:lepton} and \ref{eq:dilepton}. SM predictions including scale uncertainties are displayed in gray.}\label{fig:exp}
\end{figure}

The results of asymmetry measurements at the Tevatron and LHC experiments are summarized in Figure~\ref{fig:exp} \cite{Sha:2013} and discussed in detail in the contribution of Viatcheslav Sharyy in these proceedings. Here it shall suffice to mention the observation of an excess in $A_C^y$ at the Tevatron, while measurements of $A_C^{|y|}$ at the LHC are consistent with their standard-model (SM) prediction (and with zero) within uncertainties.

%%%%%%%%%%%%%%%%%%%%%%%%%%%%%%%%%%%%%%%%%%
%%%%%%%%%%%%%%%%%%%%%%%%%%%%%%%%%%%%%%%%%%

This write-up covers the current theoretical status of the SM prediction for the charge asymmetry (Section~\ref{sec:sm}), as well as potential contributions of new physics (Section~\ref{sec:np}). I discuss the limitations to observe $A_C^{|y|}$ at the LHC and suggest new observables involving an additional hard jet as an alternative way of measuring the charge asymmetry in proton-proton collisions (Section~\ref{sec:lhc}). I conclude in Section~\ref{sec:out} with an outlook and comments on related observables that allow a more complete picture of the charge asymmetry.

\section{Charge asymmetry in the standard model}\label{sec:sm}
\begin{figure}[tb]
\centerline{\includegraphics[width=0.85\textwidth]{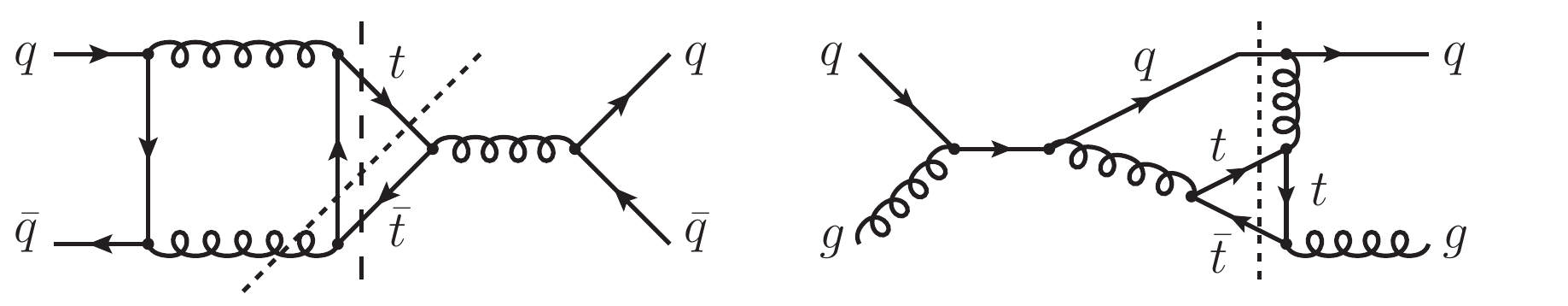}}
\caption{Charge asymmetry in QCD from quark-antiquark annihilation (left) and quark excitation (right). Shown are representative diagrams for inclusive $t\bar t$ production (dashed and dotted cuts, $qg$ contribution negligible) and $t\bar t + j$ production (dotted cuts).}\label{fig:qcd}
\end{figure}

In QCD, the charge asymmetry is generated at next-to-leading order (NLO) by additional virtual and real gluon radiation \cite{Kuhn:1998kw}, as illustrated in Figure~\ref{fig:qcd}. Normalized to the symmetric cross section, the perturbative expansion of the charge asymmetry reads
\begin{eqnarray}
A_C^{\text{QCD}} = \frac{\alpha_s^3\,\sigma_A^{(1)} + \alpha_s^4\,\sigma_A^{(2)} + \dots}{\alpha_s^2\,\sigma_S^{(0)} + \alpha_s^3\,\sigma_S^{(1)} + \alpha_s^4\,\sigma_S^{(2)} + \dots}\,.
\end{eqnarray}
%%%%%%%%%%%%%%%%%%%%%%%%%%%%%%%%%%%%%%%%%%
%%%%%%%%%%%%%%%%%%%%%%%%%%%%%%%%%%%%%%%%%%
Currently, the charge-asymmetric piece is known at NLO QCD ($\sigma_A^{(1)}$), whereas the symmetric cross section has recently been calculated up to NNLO ($\sigma_S^{(2)}$) \cite{Czakon:2013goa}. The strong remnant dependence on the factorization and renormalization scales has been significantly reduced by the resummation of large logarithms close to the partonic threshold \cite{Almeida:2008ug,Kidonakis:2011zn,Ahrens:2011uf}. The leading contribution to $A_C^{\text{QCD}}$ is dominated by the lowest-order expansion of the threshold-resummed cross section, and the asymmetry proves stable under higher-order threshold corrections. The numerical impact of fixed-order NNLO contributions ($\sigma_A^{(2)}$) on the asymmetry is not known to date, but is an important ingredient for a precise prediction of $A_C^{\text{QCD}}$.

Electroweak (EW) contributions to the charge asymmetry turn out to be significant. Fixed-order EW corrections increase the Tevatron asymmetry $A_C^y$ by about $23\%$ \cite{Hollik:2011ps}. Their effect on the LHC asymmetry $A_C^{|y|}$ is smaller due to the different parton distributions in the initial state. The resummation of EW Sudakov logarithms yields an additional enhancement of $5\%$ (apart from a minor double-counting with fixed-order corrections) \cite{Manohar:2012rs}.\footnote{Notice that EW Sudakov logarithms significantly reduce the invariant mass spectrum in $t\bar t$ production, $\text{d}\sigma_{t\bar t}/\text{d}M_{t\bar t}$, which affects constraints on potential new-physics contributions to $A_C$.} Including the leading QCD and EW fixed-order contributions, the SM predictions for the asymmetries at the Tevatron and the LHC are given by \cite{Bernreuther:2012sx}
\begin{eqnarray}
A_C^y(1.96\,\text{TeV}) = 8.75^{\,+0.58}_{\,-0.48}\,\%\,,\qquad\qquad A_C^{|y|}(7\,\text{TeV}) = 1.23\pm 0.05\,\%\,,
\end{eqnarray}
where the errors are scale uncertainties. Notice that $A_C$ decreases, if higher-order QCD corrections to $\sigma_S$ are included. This approach presumably underestimates the charge asymmetry, due to an incomplete cancellation of higher-order effects affecting both $\sigma_S$ and $\sigma_A$.

Since the results of charge asymmetry measurements are compared to predictions from Monte Carlo event generators, a precise understanding of their features is crucial for a correct interpretation. State-of-the-art Monte Carlo generators such as SHERPA and HERWIG++ with NLO matching to parton showers reproduce the qualitative features of the charge asymmetry in QCD: a decline with increasing $t\bar t$ transverse momentum $p_T^{t\bar t}$, as well as an increase with $M_{t\bar t}$ and $\Delta y$ \cite{Skands:2012mm,Hoeche:2013mua}. However, the substantial dependence of Monte Carlo predictions on the functional scale in the hard process indicates that the observed excess of the asymmetry at high $M_{t\bar t}$ and $\Delta y$ may be due to higher-order QCD and EW corrections not taken into account by Monte Carlo generators.

\section{Potential contributions from new physics}\label{sec:np}
\begin{figure}[tb]
\centerline{\includegraphics[width=1\textwidth]{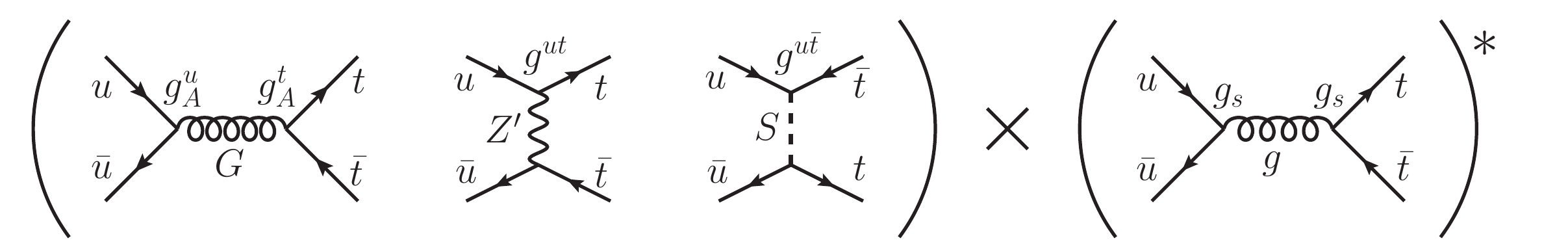}}
\caption{New-physics contributions to the charge asymmetry at tree level.}\label{fig:np}
\end{figure}

Beyond the SM, a charge asymmetry can be generated at tree level by the interference of a new $q\bar q\rightarrow t\bar t$ process with the QCD amplitude, as illustrated in Figure~\ref{fig:np}. Light new particles can generate an asymmetry as well through self-interference, if their quantum numbers prohibit an interference with the SM amplitude.
%%%%%%%%%%%%%%%%%%%%%%%%%%%%%%%%%%%%%%%%%%
%%%%%%%%%%%%%%%%%%%%%%%%%%%%%%%%%%%%%%%%%%
Possible contributions can be classified into three kinematic categories: a massive color octet with axial-vector couplings to quarks in the s-channel, a vector boson with flavor-changing couplings in the t-channel, or a new scalar in the u-channel. Such new particles and their embedding into specific models have been studied in great detail and found to be strongly constrained by correlated effects on charge-symmetric observables. In particular, the asymmetry excess at the Tevatron has stimulated a large effort to test possible new contributions at the LHC, with beneficial effects on general new-physics searches.

Among color octets with axial-vector couplings to quarks, dubbed ``axigluons'', two species yield a positive contribution to $A_C$: light axigluons ($M_G \lesssim 2 m_t$) with flavor-universal couplings, $g_A^q\cdot g_A^t > 0$ \cite{Tavares:2011zg}, or heavy axigluons ($M_G \gtrsim 2 m_t$) with opposite-sign couplings, $g_A^q\cdot g_A^t < 0$. Axigluons are predicted by a variety of models, such as chiral color featuring an extended strong gauge group \cite{Frampton:1987dn} or as Kaluza-Klein excitations in the framework of extra dimensions \cite{Bauer:2010iq}. Apart from model-inherent constraints, axigluons are mostly constrained by the invariant mass spectra of $t\bar t$ and dijet production, by the LHC asymmetry $A_C^{|y|}$, as well as electroweak precision observables \cite{Haisch:2011up, Gresham:2012kv}. Light axigluons are thus required to be broad in order to hide in the $t\bar t$ and dijet distributions measured at Tevatron and LHC. They can still account for the Tevatron excess in a mass window $200\,\text{GeV} < M_G < 450\,\text{GeV}$, which may be closed by examining the tail of angular distributions   in dijet production at the LHC. Heavy axigluons are highly challenged by a recent model-independent measurement by the CMS collaboration, which confines new-physics effects in the high-energy tail of the cross section, $\sigma_{t\bar t}(M_{t\bar t} > 1\,\text{TeV})$, to less than $20\%$ \cite{Chatrchyan:2013lca}.

Contributions of scalars in the u-channel are a priori phenomenologically disfavored, since they lead to a strong Rutherford enhancement of the $t\bar t$ invariant mass spectrum. They are excluded by measurements of atomic parity violation \cite{Gresham:2012wc}.

Further asymmetry candidates are new vector bosons with masses around $300\,\text{GeV}$ and flavor-changing neutral couplings in the t-channel, often referred to as $Z'$ bosons \cite{Jung:2009jz}. Strong constraints from flavor observables require a highly non-trivial flavor structure of their couplings, confined to right-handed up and top quarks. Such structures can be arranged for by means of flavor symmetries \cite{Grinstein:2011yv}, which also protect the new bosons from inducing undesired same-sign top production. Additional strong constraints on $Z'$ candidates arise, as for the u-channel contributions, from the $t\bar t$ and dijet invariant mass distributions and from atomic parity violation. At the LHC, a kinematic angular asymmetry in associated $Z' t$ production reconciles t-channel bosons with the measured charge asymmetry $A_C^{|y|}$ \cite{Alvarez:2012ca}. Searches for the corresponding $Z't$ resonances with $Z'\rightarrow \bar{t}u$, however, rule out an explanation of the Tevatron excess unless alternative $Z'$ decay channels dominate \cite{Drobnak:2012rb}. Since many of these constraints are model-dependent, t-channel explanations of the asymmetry are not conclusively ruled out yet. However, the search for $Z'$ bosons in top-like final states at the LHC has a high exclusion potential. Along the lines described in \cite{Craig:2011an} for a $W'$ model, t-channel explanations of the Tevatron excess may be completely ruled out by scanning existing LHC event samples from top-quark analyses for $Z'$ effects.

\section{Charge asymmetry observables at the LHC}\label{sec:lhc}
Due to the smallness of $A_C^{|y|}$, achieving a high significance for a measurement of the charge asymmetry in inclusive $t\bar t$ production at the LHC is difficult. With more luminosity during the $14\,\text{TeV}$ run, the ultimate sensitivity to the asymmetry will be limited by systematic uncertainties. A dedicated study \cite{Jung:2013vpa} shows that a significance of $95\%$ may eventually be achieved, if at least $50\%$ of the systematic errors scale with the luminosity. Given these limitations, it is advisable and maybe indispensable to consider alternative strategies to measure the top charge asymmetry at the LHC.

An interesting route to be pursued is top-antitop production in association with a hard jet in the final state. In this process, the charge asymmetry is generated already at tree level by real gluon exchange (see Figure~\ref{fig:qcd}). As a first approach, the charge asymmetry can be defined analogously to $A_C^{|y|}$ in inclusive $t\bar t$ production. In QCD, this observable has been calculated up to NLO \cite{Dittmaier:2008uj,Melnikov:2010iu}. The resulting asymmetry at the LHC at $7\,\text{TeV}$ is extremely small, $A_C^{|y|} = 0.51\pm 0.09\,\%$ \cite{Alioli:2011as}. An observation of $A_C^{|y|}$ in $t\bar t + j$ production at the LHC thus seems to struggle with even greater difficulties than inclusive $t\bar t$ production, with additional experimental challenges due to the extra jet.

However, the definition of the charge asymmetry can be improved by taking the jet {ki\-ne\-ma\-tics} into account \cite{Berge:2013xsa}. Two observables of a charge asymmetry turn out to be complementary in final-state kinematics and in their sensitivity to initial parton states: The \emph{incline asymmetry} probes the charge asymmetry from quark-antiquark annihilation, whereas the \emph{energy asymmetry} is sensitive to the asymmetry from quark excitation.

%%%%%%%%%%%%%%%%%%%%%%%%%%%%%%%%%%%%%%%%%%%%
\begin{wrapfigure}{r}{0.5\textwidth}
\centerline{\includegraphics[width=0.5\textwidth]{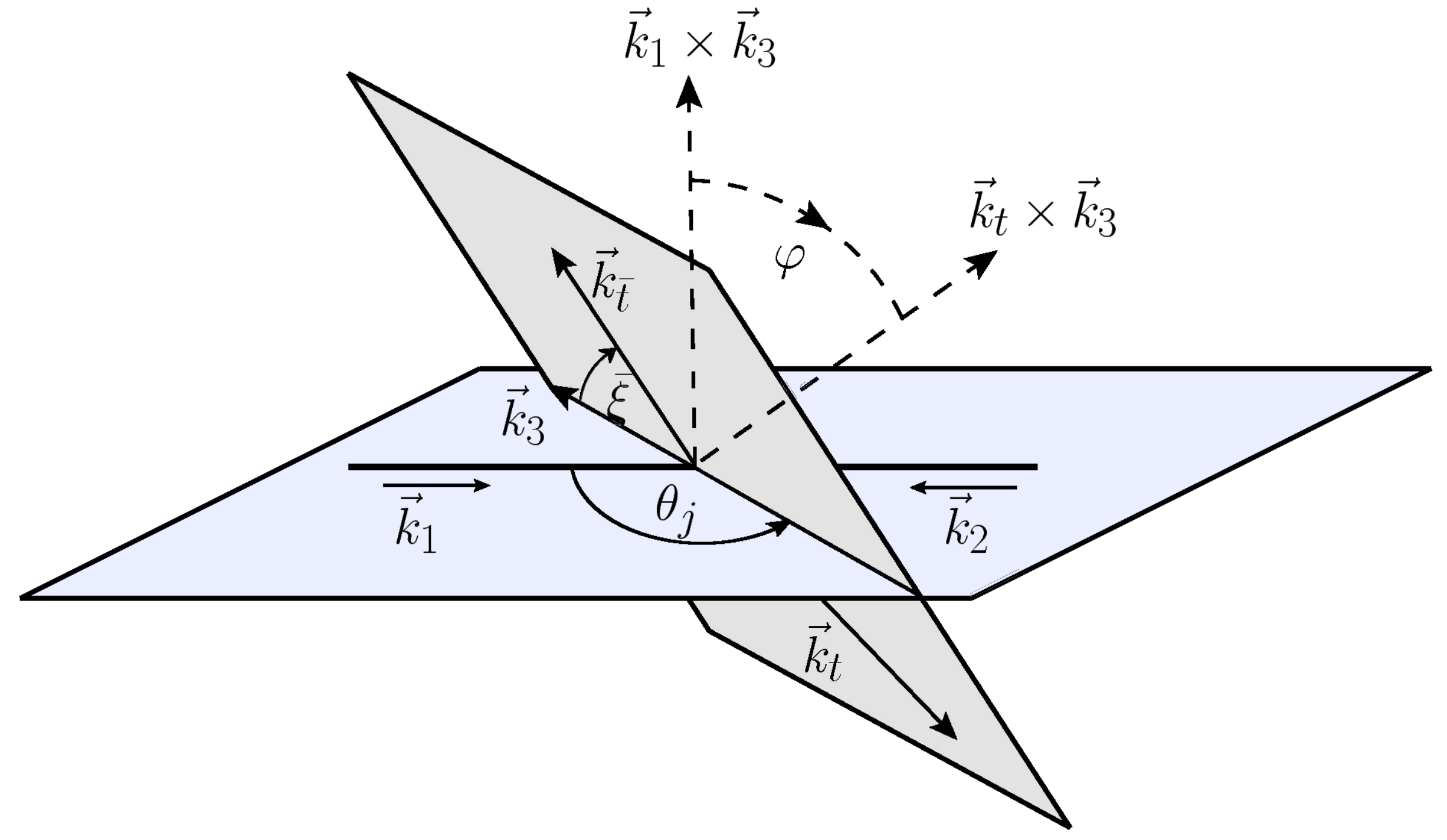}}
\vspace*{-0cm}
\caption{Kinematics for the charge asymmetry in $t \bar t + j$ production. Definition of the inclination angle $\varphi$ between the planes ($q,\bar q, j$) and ($t, \bar t, j$).}
\label{fig:incline}
\end{wrapfigure}
%%%%%%%%%%%%%%%%%%%%%%%%%%%%%%%%%%%%%%%%%%%%
The incline asymmetry is defined in terms of the inclination angle $\varphi$ between the planes spanned by the initial- and final-state quarks and the jet, as illustrated in Figure~\ref{fig:incline}. The differential incline asymmetry
\begin{eqnarray}
\frac{\text{d}\sigma_A^{\varphi}}{\text{d}\theta_j} = \frac{\text{d}\sigma(\cos\varphi > 0)}{\text{d}\theta_j} - \frac{\text{d}\sigma(\cos\varphi < 0)}{\text{d}\theta_j}
\end{eqnarray}
is largely independent of the jet scattering angle $\theta_j$ and therefore maximally sensitive to the top and antitop quarks' angular distributions. To make the incline asymmetry a {via\-ble} observable for proton-proton collisions, the direction of the incoming quark needs to be determined by focussing on boosted events with large rapidities $y_{t\bar t j}$ of the $t\bar t + j$ final state. The resulting incline asymmetry for the LHC then reads
\begin{eqnarray}
A^{\varphi}_C = \frac{\sigma_A^{\varphi}(y_{t\bar t j} > 0) - \sigma_A^{\varphi}(y_{t\bar t j} < 0)}{\sigma_S}\,.
\end{eqnarray}
With appropriate kinematic cuts, the incline asymmetry reaches up to $A_C^{\varphi} = -4\%$ at the LHC with $14\,\text{TeV}$ collision energy.

The \emph{energy asymmetry} is defined through the difference $\Delta E = E_t - E_{\bar t}$ of top and antitop energies in the parton center-of-mass frame,
\begin{eqnarray}
A^E_C & = & \frac{\sigma(\Delta E > 0) - \sigma(\Delta E < 0)}{\sigma(\Delta E > 0) + \sigma(\Delta E < 0)}\,.
\end{eqnarray}
It probes the charge asymmetry in the partonic quark-gluon channel and is equivalent to the forward-backward asymmetry of the quark-jet in the top-antitop rest frame. The energy asymmetry is well adapted to the LHC environment. It benefits from the high quark-gluon parton luminosity in proton-proton collisions and can be measured without reconstructing the direction of the incident quark. At the $14\,\text{TeV}$ LHC, the energy asymmetry reaches values of up to $A_C^E = -12\%$ in suitable regions of phase space. This new observable thus holds the potential of first observing the top-quark charge asymmetry at the LHC in $t\bar t + j$ production. As a caveat, one needs to add that the predictions for $A_C^{\varphi}$ and $A_C^E$ quoted here might be significantly changed by NLO corrections. Investigations of these contributions are underway \cite{Ber:2013}.

Another alternative measurement of the top asymmetry at the LHC has been suggested for the LHCb experiment \cite{Kagan:2011yx}. The good coverage of the forward region by the LHCb detector may allow to measure top-antitop rapidity differences in the region of large rapidities, where the charge asymmetry is maximal.

\section{Outlook and related observables}\label{sec:out}
The origin of the asymmetry excess at the Tevatron remains a puzzle. While the measurement of a charge asymmetry at the LHC is valuable on its own, the comparison with the Tevatron asymmetry will always be limited due to the different experimental conditions. To shed light on the Tevatron anomaly and to gain further insight into various models in connection with the charge asymmetry, several related observables have been proposed and in some cases been measured.

The charge asymmetry in $t\bar t$ production can be measured via the angular distributions of leptons from the top decays without reconstructing the top quarks \cite{Bowen:2005ap,Krohn:2011tw}. Two observables have been probed by experiments, the single-lepton asymmetry
\begin{eqnarray}\label{eq:lepton}
A_C^{\ell} = \frac{\sigma(q\cdot\eta_{\ell} > 0) - \sigma(q\cdot \eta_{\ell} < 0)}{\sigma(q\cdot \eta_{\ell} > 0) + \sigma(q\cdot\eta_{\ell} < 0)}\,,
\end{eqnarray}
where $q$ and $\eta_{\ell}$ are the lepton's charge and pseudo-rapidity, and the dilepton asymmetry
\begin{eqnarray}\label{eq:dilepton}
A_C^{\ell\ell} = \frac{\sigma(\Delta\eta > 0) - \sigma(\Delta\eta < 0)}{\sigma(\Delta\eta > 0) + \sigma(\Delta\eta < 0)}\,,
\end{eqnarray}
in terms of the rapidity difference $\Delta\eta = \eta_{\ell^+} - \eta_{\ell^-}$ between leptons from the top and antitop decays. The experimental results for these asymmetries are shown in Figure~\ref{fig:exp}. The relation between the lepton asymmetry and the top-antitop asymmetry is model-dependent. Lepton asymmetries thus prove particularly useful in distinguishing between models with chiral top-quark couplings \cite{Falkowski:2012cu}.

Another proposal considers a measurement the forward-backward asymmetry of bottom quarks at the Tevatron \cite{Grinstein:2013mia}. Above the $Z$ pole, the observable asymmetry is dominated by QCD contributions. Beyond the SM, the bottom charge asymmetry allows to probe the flavor structure of new-physics contributions to the top asymmetry.

%%%%%%%%%%%%%%%%%%%%%%%%%%%%%%%%%%%%%%%%%%%%%%%%
 
% ****************************************************************************
% BIBLIOGRAPHY AREA
% ****************************************************************************

% please do not change the following line
\begin{footnotesize}

% please do not change the following line
\end{footnotesize}

% ****************************************************************************
% END OF BIBLIOGRAPHY AREA
% ****************************************************************************

\end{document}